# Using Categorical Features In Mining Bug Tracking Systems to Assign Bug Reports

Mamdouh Alenezi[1], Shadi Banitaan[2], and Mohammad Zarour[1]

[1]College of Computer & Information Sciences, Prince Sultan University, Riyadh 11586, Saudi Arabia
[2]Computer Science and Software Engineering, University of Detroit Mercy, USA

## ABSTRACT

*Most bug assignment approaches utilize text classification and information retrieval techniques. These approaches use the textual contents of bug reports to build recommendation models. The textual contents of bug reports are usually of high dimension and noisy source of information. These approaches suffer from low accuracy and high computational needs. In this paper, we investigate whether using categorical fields of bug reports, such as component to which the bug belongs, are appropriate to represent bug reports instead of textual description. We build a classification model by utilizing the categorical features, as a representation, for the bug report. The experimental evaluation is conducted using three projects namely NetBeans, Freedesktop, and Firefox. We compared this approach with two machine learning based bug assignment approaches. The evaluation shows that using the textual contents of bug reports is important. In addition, it shows that the categorical features can improve the classification accuracy.*

## 1. INTRODUCTION

Software maintenance is crucial in software evolution which aims to modify a software product after delivery to fix defects or to improve other attributes. It represents one of the most expensive and time-consuming phases in the whole development process. Previous studies showed that around 90% of software development cost is utilized on maintenance and evolution activities [1]. Improving the competence of the bug fixing process would lower the costs of software development as indicated by several studies [2]. The bug fixing process would benefit greatly from improving the bug assignment accuracy by assigning bugs to appropriate developers.

Several software repositories such as source code control, archived communications, and bug repositories are available for many software projects. A software project usually sustains a bug repository (a.k.a bug tracking system (BTS)) that manages and tracks bug reports and their status. BTS allows both developers and users to submit defects, suggest enhancements, and comment on bug reports. These bug reports are then utilized to guide software maintenance activities in order to produce more robust systems. The usage of a bug repository can improve the development process in different aspects: allowing developers who are not co-located to share their knowledge about project development [3]. Tracing the project evolution [20], and improving the quality of the project [4].

Bug assignment is an essential step in bug fixing where a relevant developer is assigned to a new submitted bug. A person who performs this task is called triager.

A triager usually spends effort and time to make several triaging decisions such as finding a relevant developer to fix a newly submitted bug report and determining if a new bug report is valid or duplicate. Every decision needs heavy human involvement to collect relevant information and search through the repository. Relying on the triager knowledge and judgment, after sieving through huge repositories, makes bug assignment error-prone and time-consuming. On the same hand, the number of daily submitted bugs is vastly huge which makes the manual triaging an unproductive process. To solve these problems, some machine learning algorithms are employed to conduct automatic bug assignment [5]–[9]. Most of the bug assignment approaches are based on text categorization [5],[10]. However, these approaches suffer from low-quality bug reports which may mislead the assignment approach to assign bugs to wrong developers [7], [11]. These approaches also suffer from low recall values [5].

In this paper, we investigate whether using categorical fields is appropriate to represent bug reports instead of textual description. The main contributions of this paper are:

a. We build a recommendation model to predict a developer to a newly submitted bug report using bug reports categorical data. We investigate the effect of using several meta-features of bug reports such as the component that each bug belongs to on the classification accuracy.
b. We perform an experimental evaluation on three open source projects to investigate the effect of both the textual content and the categorical features, as bug reports representations, on the recommendation accuracy.

The paper is organized as: Section 2 introduces some background material related to this work. Section 3 discusses related work. Section 4 describes the proposed approach. The experimental evaluation and discussions are presented in section 5. Section 6 presents some threats to validity and Section 7 concludes the paper.

## 2. BACKGROUND

We create a recommendation model based on categorical data found in bug reports to assist triagers in making decisions. When a developer or a user encountered a bug while using the software or wants to request an enhancement, he/she usually open a bug report in the open bug repository. Many open bug repositories (e.g., Bugzilla, Jira, Gnats, and Trac) have been adopted in open source projects. We only explore projects that use Bugzilla as their bug tracking system.

### 2.1. BUG REPORTS

In Bugzilla, bugs are kept in the form of bug reports which consist of predefined fields, text description, and attachments. Predefined fields represent several categorical attributes of a bug report. Some attributes such as creation date are unchangeable. Other attributes may be changed over time such as product, priority, and severity. Some attributes such as the final resolution may be frequently modified by authorized persons [5]. The assignee field represents the person in charge of resolving the bug. It contains the ID of the user who was made responsible for providing a solution for a particular bug report. Figure 1 shows an example of a bug report.

Figure 1: An Example of Bug Report

### 2.2. BUG LIFE-CYCLE

Bug reports experience several states in their life-cycles. Figure 2 depicts the life-cycle of bugs in Bugzilla. When a new bug report is submitted, its state is set to NEW. Once it has been triaged and assigned to a developer, its state is then changed to ASSIGNED. If this bug has been closed, its state is set to RESOLVED, VERIFIED or CLOSED depending on its situation. A report can be resolved in several ways; the resolution status in the report is used to record how the report was resolved. If the resolution results in changing the code base, then it is marked as FIXED. When a bug is determined as duplicated to other bugs, it is set to DUPLICATE. If a bug will not be fixed, or it is not an actual bug, it will be set to WONTFIX or INVALID respectively. If a bug was once resolved but has been reopened, it is marked as REOPENED [5].

### 2.3. Classifications

Classification is one of the most commonly used machine learning techniques. It is also known as supervised statistical learning. In supervised learning, the model needs to be first trained using data with predetermined classes. This data is used to train the learning algorithm, which creates a model that can then be used to label/classify the testing instances, where the values of the class labels are unknown. Classification techniques have been successfully used in many fields such as computer vision, speech recognition, natural language processing, and document classification.

One of the popular classification techniques is the Naive Bayes classifier [12]. It is a probabilistic classifier based on applying Bayes' theorem. It assumes that the presence or absence of a particular feature is unrelated to the presence or absence of any other features, given the class variable.

A classification problem can be written as the problem of finding the class with maximum probability given a set of observed features values. The posterior probability of the class is computed using the Bayes theorem, as:

$$P(C|X) = \frac{P(X|C)P(C)}{P(X)}$$

where X is features, X = {$x_1$; $x_2$, ..., $x_m$} and C is a Class, P(C) is the prior probability of the class and P(X|C) is the conditional probability of the features given the class. P(X|C) is computed as follows:

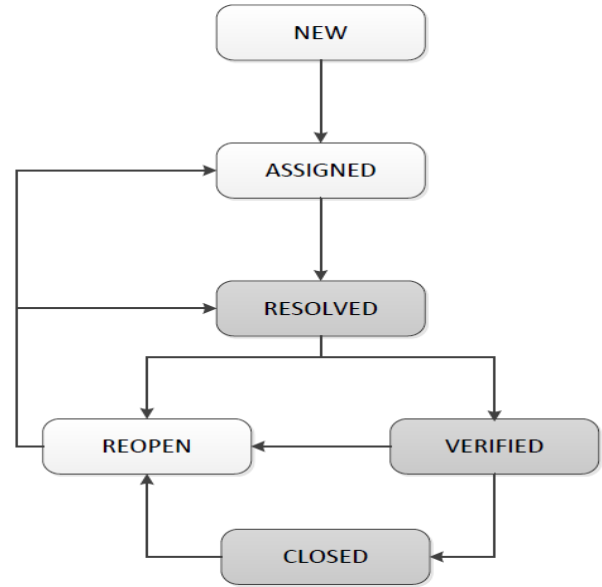

Figure 2: Bug Life Cycle

$$P(X|C) = P(x1, x2, \ldots, xm|C) = \prod_{j=1}^{m} P(xj|C)$$

where m is the number of features and the highest P(C|X) is predicted as the class label of X. Naive Bayes is one of the popular classifiers in terms of accuracy and computational efficiency [13].

## 3. RELATED WORK

Machine learning techniques have been applied on bug tracking systems to detect duplicate bug reports [1], estimate the required effort to fix bug reports [14], predict the priority of the reported bug [15], and predict the files that have most post-release defects [16]. Machine learning based recommender is one of the five categories proposed in [17] to the bug assignment problem.
Most bug assignment approaches utilize text classification and information retrieval techniques. In this work, we categorize bug assignment approaches into two categories. The first category assigns only one developer to a newly submitted bug report [5], [18]–[20]. The second category assigns a set of candidate developers to a newly submitted bug report [6], [21]–[23].

Approaches in the first category formulate the bug assignment problem as a classification task. A model is built on the training dataset and then used to predict a developer for each new testing instance. For training, information retrieval techniques were used to represent bug reports. A class label is assigned to each bug report in the training data using the assignee field of that bug report. Using Bayesian algorithm is proposed in [18] to train a classifier with the textual content of resolved bug reports. Then this classifier is used to classify newly incoming bug reports. They predicted 30% of the report assignments correctly, considering Eclipse as a case study. Anvik et al. [5] improved the approach proposed in [18] by removing inactive developers (i.e, developers with a too low bug fixing frequency or developers whom not working on the project anymore) and using project-specific heuristics to label bug reports. They employed SVM, Naive Bayes and Decision Trees classification techniques, and reported prediction accuracy of up to 64%. Zou et al. [19] proposed the training set reduction with both feature selection and instance selection techniques for bug assignment. They evaluated their approach on Eclipse where their approach removed 70% words and 50% bug reports. Alenezi et al. [20] employed five state-of-the-art term selection methods namely Log Odds Ratio, Chi-Square, Term Frequency Relevance Frequency, Mutual Information, and Distinguishing Feature Selector on the textual description of bug reports to produce discriminating terms. After that, they built a classification model on the discriminating terms using Naive Bayes classifier. Their experimental results on four real datasets showed that by selecting a small number of discriminating terms, the classification accuracy can be significantly improved. Banitaan and Alenezi [24] enhanced the classification accuracy by utilizing the most discriminating terms of bug reports based on Chai-square, the components in which the bugs belong to, and the reporter who filed the bug. They compared their approach with Anvik et al. [5] and Alenezi et al. [20] then reported an improvement in the classification accuracy.

Bug assignment in the second category considered the bug fixing process as a collaborative effort. Even though a bug report is assigned to only one developer, many developers contribute their expertise which helped in solving it. Matter et al. [6] modeled a developer's expertise using the vocabulary found in the developer's source code. They recommended experienced developers by extracting information from new bug reports and looking it up in the vocabulary. Their approach was tested on 130,769 Eclipse bug reports and achieved 33:6% top-one precision and 71% top-ten recall using eight years of Eclipse project. Wu et al. presented a bug assignment approach called DREX [21]. For each newly submitted bug report, DREX used K-Nearest-Neighbour to search for similar bug reports and then built a social network of their developers and commenters. Based on several social network metrics, they ranked these developers to recommend them to the new bug report. Their experimental evaluation on Firefox showed that the two metrics (Out-degree and Frequency) gave the best performance. Xie et al. [25] proposed an approach that models developers' expertise based on topic modeling. Their approach recommends a ranked list of developers who are candidates for resolving the new bug. Their experimental results on Eclipse JDT and Mozilla Firefox projects showed that their approach achieves high recall values. Zhang and Lee [23] proposed a developer recommendation technique to recommend candidate developers to fix new bugs. They used both social network and experts' feedback to recommend potential developers. They considered both the experience of these potential developers and the fixing efficiency to rank them. Their approach achieved F-score of 40% when recommending 3 developers. Similarly, Kempe et. al. [25] used social networks for the same purposes. Tamrawi et al.[26], on the other hand, employed fuzzy sets to model bug-fixing expertise of developers such that developers who recently fixed bugs are more likely to fix similar bugs in the near future. They used some recent reports to build the fuzzy-sets representing the membership of developers to technical terms in the reports. For newly reported bugs, developers are recommended by comparing their membership to the terms included in the new report. Banitaan and Alenezi [27] built developers social networks based on developers' comments on bug reports. They used the detected developers' communities to assign a relevant community to each newly committed bug report.

Our approach belongs to the first category in which they assign only one developer for each newly submitted bug. Different from previous work which use the unstructured textual content of bug reports to represent them, our approach utilizes structural data bug reports to represent them.

## 4. THE APPROACH

We present an approach for the automatic recommendation of one experienced developer for each new bug report. Our approach uses a machine learning algorithm to recommend a developer who may be appropriate for resolving the bug. Adopted from previous work, we formulate the bug triaging process as a classification task where instances represent bug reports, features represent different meta-data fields of the report, and the class label represents the developer who fixed this report.

Previous bug assignment approaches utilized the textual form of bug reports to build their models. Building recommendation models based on the textual description of bug reports suffer from several limitations. The textual content of bug reports leads to a very large corpus with so many sparse terms to describe bug reports. It is also a very noisy source of data and needs to go through different pre-processing steps. In addition, users and developers change over time which leads to different vocabulary usage to describe similar issues [28]. The aforementioned limitations lead to computational expensive and poor accuracy results.

In classification, features do not contribute equally to identify class labels. In addition, many features may be irrelevant (i.e., they do not contribute to classification). Including irrelevant features leads to the high dimensionality of data and poor accuracy results. Therefore, we should select only the important features that can distinguish between different class labels (developers). This selection may result in enhanced performance.

In our work, we use gain ratio [12] to identify the important features that have a high contribution to classification. The gain ratio provides a normalized measure of the contribution of each feature to classification. After applying the gain ratio to all meta-data fields, we find that Component, Operating System, and Priority have a high gain ratio value while other fields such as Severity are irrelevant (i.e., they have low gain ratio values). Some other fields such as Keywords cannot be used as features in classification since they are optional (i.e., most bug reports have empty values for these fields).

In our approach, we use the following features:

a. Component: It represents the component to which the bug belongs. For example, in the Netbeans dataset, some of the components are Java, Compiler, UI, and Ant.

b. Operating System: It represents the operating system against which the bug was reported. For example, Windows 7, Mac OS X, and Linux.

c. Priority: It represents the importance and order in which a bug should be fixed compared to other bugs. There are five priority levels P1, P2, …, P5 where P1 is considered the highest and P5 is the lowest.

## 5. EXPERIMENT EVALUATION

In the Section, we present our experimental evaluation. Section 5.1 presents the datasets used in this work, Section 5.2 presents the results of comparing the new approach with two other approaches.

## 5.1 DATASETS

Three datasets namely NetBeans[1], Freedesktop[2], and Firefox[3] are used to evaluate the approaches. NetBeans is an integrated development environment (IDE) for developing primarily with Java, but also with other languages, in particular, PHP, C/C++, and HTML5. Free desktop consists of different open source software projects working on interoperability and shared technology for X Window System desktops. Firefox is a widely used free and open source web browser. It is developed for Windows, OS X, Linux, and Android by Mozilla Foundation. All bug reports and their data are available and downloaded from the bug tracking systems of the corresponding projects.

We collected bug records that have the status of [Closed, Verified, and Resolved] and the resolution of [Fixed]. For NetBeans, we choose 9760 bug reports from December 1st, 2011 until December 31st, 2012. For Free desktop, we choose 5817 bug reports from December 1st, 2011 until December 31st, 2012. For Firefox, we choose 3132 bug reports from December 1st, 2011 until December 31st, 2012. We want to refine the training set further to remove reports that are assigned to inactive developers (i.e., developers who no longer work on the project or developers who have only fixed a small number of bugs). We only consider developers who have fixed at least 15 bug reports in the dataset. Table 1 shows a summary of the refined datasets.

Table 1. Statistics of all Bug Reports Data

| Name | # of Bug Reports | # of Developers |
|---|---|---|
| NetBeans | 9500 | 58 |
| Freedesktop | 5129 | 61 |
| Firefox | 2488 | 40 |

## 5.2 RESULTS

We use the Naive Bayes classifier in the proposed approach and we recommend one developer for each new bug report. For evaluation, the dataset is divided into training and testing sets. To obtain unbiased evaluation results, we perform a 10-fold cross-validation. Precision, Recall, F-score, and AUC are used for evaluation.

$$Precision = \frac{\text{\# of correct recommendations}}{\text{\# of recommendations made}}$$

$$Recall = \frac{\text{\# of correct recommendations}}{\text{\# of possible relevant developers}}$$

$$F - Score = 2 \times \frac{\text{Precision x Recall}}{\text{Precision + Recall}}$$

The area under the receiver operating characteristic curve statistic (AUC) is a robust measure to assess and compare the performance of classifiers since it is independent of prior probabilities. In order to investigate the effect of using meta-data features other than the textual content of bug report on the classification accuracy, we compare the approach that utilizes the categorical features only with two different approaches. We will refer to proposed approach by CF. The first

---

[1] http://netbeans.org/bugzilla/

[2] https://bugs.freedesktop.org/

[3] https://bugzilla.mozilla.org/

International Journal of Software Engineering & Applications (IJSEA), Vol.9, No.2, March 2018approach represents bug reports using the textual contents only [20]. We will refer to this approach by CHI-2. The CHI-2 approach works as follows:

1. The summary of bug reports is used as a description of bugs.
2. The traditional text processing is applied to remove white-spaces, punctuation, numbers, and stopwords.
3. A bug-term matrix is constructed and weighted by term frequency.
4. The $X^2$ term selection method is used to select the best 10% distinctive terms.

The second approach, TRAM represents bug reports by using both the textual contents and some categorical features [24]. The following features are used in the TRAM approach:

1. The 1% most discriminating terms of the textual description of bug reports selected by $X^2$ method.
2. Component to which the bug belongs to.
3. The reporter who filed the bug report.

Table 2 presents the classification results of the three approaches for the selected projects. It is clear that TRAM outperforms both CF and CHI2 in all datasets in terms of Precision, Recall, F-Score, and AUC. It improves the F-Score over the CF by 1:5%, 8:6%, and 15:5% for NetBeans, Freedesktop, and Firefox respectively. TRAM also improves the F-Score over the CHI2 by 35:8%, 27:9%, and 17:5% for NetBeans, Freedesktop, and Firefox respectively. We can also notice that CF outperforms CHI2 by 34:3%, 19:3%, and 2:2% for NetBeans, Freedesktop, and Firefox respectively. Blending textual and categorical features gave the best classification results. In addition, using the categorical features only gave good results and can be used to reduce the dimensionality of features. The best F-Score achieved by TRAM in the Freedesktop project is (67%). Since none of the approaches achieved high accuracy, we believe that classification is not the best approach to solve bug assignment problem. Other approaches could be investigated such as Recommender Systems and Swarm Intelligence.

Table 2.  Classification Results

| Project | Method | Precision | Recall | F-Score | AUC |
|---|---|---|---|---|---|
| NetBeans | CHI2 | 0.38 | 0.234 | 0.290 | 0.737 |
|  | TRAM | 0.663 | 0.634 | 0.648 | 0.957 |
|  | CF | 0.652 | 0.615 | 0.633 | 0.947 |
| Freedesktop | CHI2 | 0.423 | 0.363 | 0.391 | 0.80 |
|  | TRAM | 0.685 | 0.656 | 0.670 | 0.950 |
|  | CF | 0.6 | 0.569 | 0.584 | 0.943 |
| Firefox | CHI2 | 0.361 | 0.355 | 0.358 | 0.780 |
|  | TRAM | 0.537 | 0.533 | 0.535 | 0.898 |
|  | CF | 0.373 | 0.388 | 0.380 | 0.859 |

## 6. THREATS AND VALIDITY

In this Section, we enumerate some threats to validity in our work. Although we considered the developer in the assignee field as the developer who fixed that bug, bug fixing can be seen as a collaborative effort where several developers collaborate to fix a bug. Therefore, using developer's collaborative effort can lead to recommending several developers who have a potential experience in solving a bug. In our study, we used three large open source projects, NetBeans, Freedesktop, and Firefox.

These projects have multiple components and operating systems; hence we could use this information as features for our recommendation models. For comparatively smaller projects which do not have components or operating systems, the lack of component- operating System labels would reduce accuracy. Threats to external validity concern the generalization of our findings. Every result obtained through empirical studies is threatened by the bias of their datasets. We have validated our approaches using open source bug repositories only; closed source projects may have different properties that may require some modifications. Moreover, we only use projects that use Bugzilla bug tracking system. Other bug tracking systems are available such as JIRA and TRAC that model bug reports differently. Therefore, our approaches should be applied to the more open source and commercial projects in order to generalize their results.

## 7. CONCLUSIONS

In this paper, we have investigated whether using categorical fields of bug reports is better than using the textual contents to represent bug reports in the recommendation model. The categorical fields of bug reports were used to represent bug reports. A recommendation model was built based on a training data of bug reports history and then it is used to predict a developer with a newly coming bug report. We performed a 10-fold cross validation to get unbiased results. Our experimental evaluation on three projects showed that using textual and categorical features together gave better classification results. Since none of the approaches achieved high accuracy, we believe that classification is not the best approach to solve bug assignment problem. Other approaches could be investigated such as Recommender Systems and Swarm Intelligence. In the future, we are planning to deploy the recommendation model in an industrial setting to evaluate its practical usage.

**AUTHORS**

**Dr. Mamdouh Alenezi** is currently the Chief Information & Technology Officer (CITO) at Prince Sultan University. Dr. Alenezi received his MS and PhD degrees from DePaul University and North Dakota State University in 2011 and 2014, respectively. He is a member of the Institute of Electrical and Electronic Engineers (IEEE). His is research interests include software engineering, open source software, software security and data mining. He teach mainly software engineering courses.

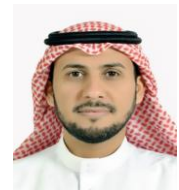

**Dr. Banitaan** is currently the director and associate professor at the Mathematics, Computer Science, and Software Engineering department at the University of Detroit Mercy. His research interests include software engineering and data mining. He is a member of the Association for Computing Machinery (ACM), a member of the Institute of Electrical and Electronic Engineers (IEEE), and a member of the IEEE Computer Society. He received a B.S. degree in Computer Science from Yarmouk University, an M.S. degree in Computer and Information Sciences from Yarmouk University, and a

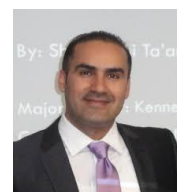

Ph.D. degree in Computer Science from North Dakota State University. He taught for five years at the University of Nizwa, Oman. He joined the University of Detroit Mercy in 2013.

**Dr. Zarour** holds a Ph.D. in Software Engineering (2009) from École de Technologie Supérieure (ETS) – Université du Québec (Montréal, Canada) and master degree in Computer Science (1998) from University of Jordan. He is currently a faculty member in college of computer and information sciences (CCIS) at Prince Sultan University, Riyadh, Saudi Arabia. He has more than 12 years of teaching experience in university and academic environment and also has several years of industry experience in information systems development and project management. His research interests include software process assessment and improvement, software quality, cost estimation and web technologies. He has many peer-reviewed publications. 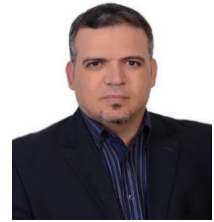